\documentclass{aastex63}

\usepackage{graphicx}
\usepackage{multirow}
\usepackage{tabularx}
\usepackage{longtable,tabu}
\usepackage{subfigure}
\usepackage{amsmath}
\usepackage{ulem}

\received{}
\revised{}
\accepted{}
\submitjournal{PASP}
\shorttitle{CONTROL}
\shortauthors{Sreejith et al.}

\begin{document}

\title{THE AUTONOMOUS DATA REDUCTION PIPELINE FOR THE CUTE MISSION}
\correspondingauthor{A. G. Sreejith}
\email{sreejith.aickara@oeaw.ac.at}

\author{A. G. Sreejith}
\affiliation{Space Research Institute, Austrian Academy of Sciences, Schmiedlstrasse 6, 8042 Graz, Austria}
\affiliation{Laboratory for Atmospheric and Space Physics, University of Colorado, Boulder, CO, 80303, USA}
\author{Luca Fossati}
\affiliation{Space Research Institute, Austrian Academy of Sciences, Schmiedlstrasse 6, 8042 Graz, Austria}
\author{S. Ambily}
\affiliation{Laboratory for Atmospheric and Space Physics, University of Colorado, Boulder, CO, 80303, USA}
\author{Arika Egan}
\affiliation{Laboratory for Atmospheric and Space Physics, University of Colorado, Boulder, CO, 80303, USA}
\author{Nicholas Nell}
\affiliation{Laboratory for Atmospheric and Space Physics, University of Colorado, Boulder, CO, 80303, USA}
\author{Kevin France}
\affiliation{Laboratory for Atmospheric and Space Physics, University of Colorado, Boulder, CO, 80303, USA}
\author{Brian T. Fleming}
\affiliation{Laboratory for Atmospheric and Space Physics, University of Colorado, Boulder, CO, 80303, USA}
\author{Stephanie Haas}
\affiliation{Space Research Institute, Austrian Academy of Sciences, Schmiedlstrasse 6, 8042 Graz, Austria}
\author{Michael Chambliss}
\affiliation{Laboratory for Atmospheric and Space Physics, University of Colorado, Boulder, CO, 80303, USA}
\author{Nicholas DeCicco}
\affiliation{Laboratory for Atmospheric and Space Physics, University of Colorado, Boulder, CO, 80303, USA}
\author{Manfred Steller}
\affiliation{Space Research Institute, Austrian Academy of Sciences, Schmiedlstrasse 6, 8042 Graz, Austria}

\begin{abstract}
The Colorado Ultraviolet Transit Experiment (CUTE) is a 6U NASA CubeSat carrying on-board a low-resolution, near-ultraviolet (2479--3306\,\AA) spectrograph. It has a  Cassegrain telescope with a rectangular primary to maximize the collecting area, given the shape of the satellite bus, and an aberration correcting grating to improve the image quality, and thus spectral resolution. CUTE, launched on the 27th of September 2021 to a Low Earth Orbit, is designed to monitor transiting extra-solar planets orbiting bright, nearby stars to improve our understanding of planet atmospheric escape and star-planet interaction processes. We present here the CUTE autONomous daTa ReductiOn pipeLine (CONTROL), developed for reducing CUTE data. The pipeline has been structured with a modular approach, which also considers scalability and adaptability to other missions carrying on-board a long-slit spectrograph. The CUTE data simulator has been used to generate synthetic observations used for developing and testing the pipeline functionalities. The pipeline has been tested and updated employing flight data obtained during commissioning and initial science operations of the mission.
\end{abstract}


\section{Introduction}\label{sec:Introduction}
Transmission spectroscopy of transiting exoplanets provides several clues into the structure and composition of their atmospheres, which can then be used to constrain models of planetary atmospheric structure, composition, and evolution. In this respect, short-period exoplanets are particularly useful for several reasons. They are hotter than longer period ones, which increases the atmospheric pressure scale height, hence the atmospheric signal probed by transmission spectroscopy. Furthermore, several transits can be observed within a short amount of time. The Colorado Ultraviolet Transit Experiment \citep[CUTE;][]{fleming,france2020,Egan2020,kevin2022,Egan2022} is a 6U form factor CubeSat designed to perform near-ultraviolet (NUV) exoplanet transmission spectroscopy to characterise planetary upper atmospheres aiming at constraining atmospheric escape, which strongly affects the long-term atmospheric evolution of a planet. CUTE was launched on September 27th 2021 into a Sun-synchronous orbit at an altitude of about 570\,km. CUTE data are relayed onto the ground station located at the Laboratory for Atmospheric and Space Physics (LASP, Boulder, Colorado, USA), and are going to be archived at the NASA Exoplanet Archive. 

While observing, CUTE produces a spectrum roughly every five minutes, which calls for a fast, reliable, and automatic data reduction pipeline. The pipeline shall also perform all the necessary checks to ensure a correct execution and to recognise possible instrumental problems. The pipeline shall also be modular to enable the user to intervene at any reduction step, possibly apply the necessary updates, and thus enable running the different reduction steps separately. 

We present here the CUTE data reduction pipeline called CONTROL (CUTE autONomous daTa ReductiOn pipeLine) and its calibrated data products. The pipeline is designed to run completely autonomously with the help of a parameter file or default values (see Section~\ref{sec:control_gd}) to produce science quality master calibration files and wavelength- and flux-calibrated spectra for each science exposure. The version of the pipeline described here was designed before launch and tested extensively based on simulated data produced by the CUTE data simulator \citep{cutedrndl}. The pipeline also incorporates modifications implemented as the result of the analysis of data obtained during spacecraft and payload commissioning and the early phases of science observations, and development will continue as the mission progresses\footnote{The user shall refer to the pipeline web page for future updates {\it http://lasp.colorado.edu/home/cute/}}. In Section \ref{sec:cute}, we present a brief summary of the CUTE mission and instrument. Section \ref{sec:control_overview} gives an overview of the pipeline and describes the data levels. Section \ref{sec:architecture} presents the detailed architecture and functionality of the pipeline. Our conclusions are drescribed in Section \ref{sec:conclusion}. The appendix contains a detailed description of the various file systems used in the pipeline and the complete header keywords.
\section{The CUTE Instrument}\label{sec:cute}
CUTE is a 6U form factor CubeSat in which 4U are used for the scientific payload and 2U for payload and satellite electronics, including avionics, and communication. The CUTE payload consists of a rectangular Cassegrain telescope feeding light into a low-resolution spectrograph with an ion etched, aberration correcting, holographicaly ruled grating operating from 2479\,\AA\ to 3306\,\AA\ through a long slit. The rectangular shape of the primary enables a factor of three improvement in collecting area as compared to a standard circular aperture. A detailed description of CUTE design, hardware, and operations is given by \citet{fleming} , \citet{egan} and \citet{nell2021}. CUTE's aperture, wavelength coverage, and spectral resolution enable the detection of atomic species such as Fe{\sc i}, Fe{\sc ii}, Mg{\sc i}, and Mg{\sc ii} lines present at NUV wavelengths in the upper atmospheres of close-in giant exoplanets. The spectrum is imaged onto a Teledyne e2v CCD42-10 back-illuminated, UV-enhanced CCD detector with an active area of 2048$\times$512 pixels and with a pixel size of 13.5 $\mu$m. The CCD42-10 has flight heritage, as the sensor was used on the Mars Science Laboratory ChemCham LIBS spectrometer. 
The spacecraft bus including avionics and attitude control is provided by Blue Canyon Technologies (BCT) and can provide attitude control to better than 7 arcseconds along each of the three control axes.  
 
CUTE was launched on the 27th of September 2021 into a Sun-synchronous orbit at an altitude of about 570\,km, corresponding to a satellite orbital period of about 90 minutes. The CubeSat is operated from the Mission Operations Center (MOC) located at LASP, which has been successfully used for other LASP CubeSats.  In routine science operations, each CUTE exposure is of 300 seconds. CUTE observes each transit for a minimum of five times the transit duration to ensure a well-established stellar flux baseline and coverage of any time-variable transit event. Considering an average planetary orbital period of three days, CUTE can observe up to 3 transits per week. Each target is being visited for multiple transits (i.e. $\geq6$) to ensure complete transit coverage and to search for variability in both transit shape and depth. 

CUTE employs an S-band antenna to send down the data to the ground station located in Boulder and a UHF antenna for basic up- and down-link communication and as a backup down-link for the data in case of failure of the S-band antenna. To reduce the amount of data to be downloaded from the spacecraft, each full-frame CUTE image is trimmed on-board such that only 50 pixels above and below the spectrum centroid are sent to MOC. The downloaded data are then reduced on the ground. In the event that the UHF transmitter becomes the primary down-link channel, we will employ on-board processing that is also described in Section 4. In this case, we will transmit the reduced and extracted one-dimensional spectra, cross dispersion profiles, housekeeping data, and several full-frame images a week to verify on-board processing and target centering.  
\section{Overview of CONTROL}\label{sec:control_overview}
\begin{figure}[h!]
\begin{center}
\includegraphics[width=\textwidth,height=\textheight,keepaspectratio]{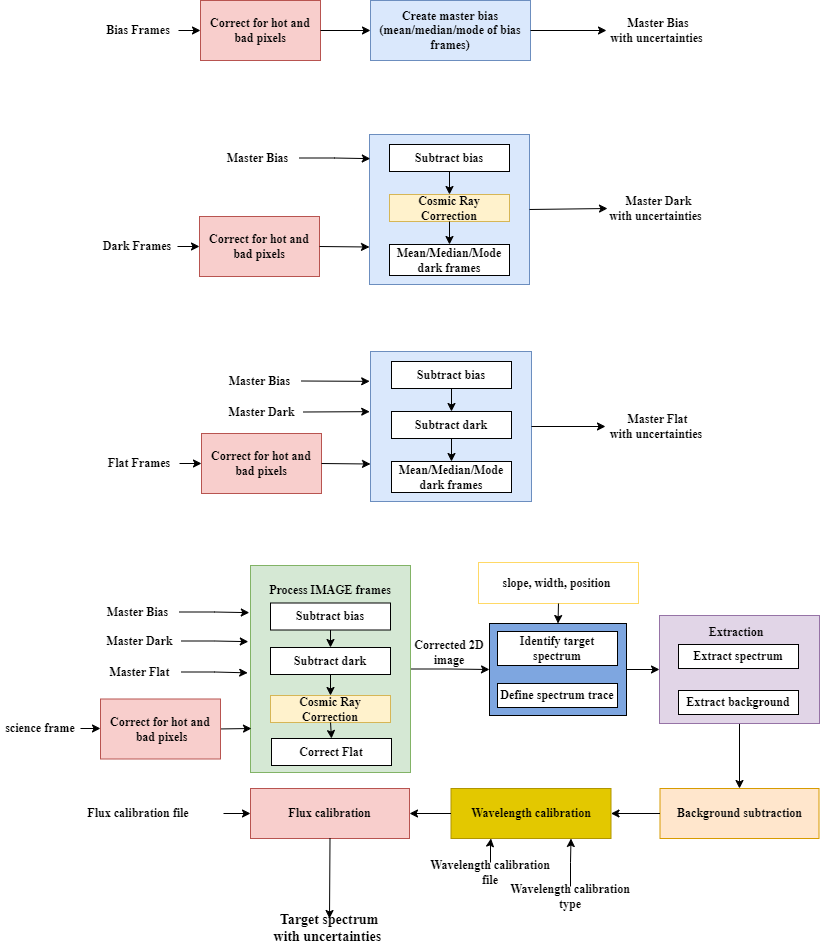}
\caption{ Data reduction flowchart for complete operations capable by CONTROL. The hot and bad pixel modules are shaded in red, the calibration master files modules in  light blue, the cosmic ray removal module in light orange, the spectrum identification module in grey, the spectrum trace and extraction modules in violet, the background subtraction module in orange, the wavelength calibration module in yellow, and the flux calibration module in dark blue.}
\label{fig:fig1}
\end{center}
\end{figure}

CONTROL provides end-to-end data reduction of CUTE data without human intervention and supervision. CONTROL can perform dark and bias subtraction, flat-field correction, corrections for bad/hot pixels, cosmic-ray correction, spectral extraction, background subtraction, wavelength calibration, and flux calibration as required by the user. The data reduction pipeline is therefore also able to combine dark, bias, and flat frames to produce master calibration frames. On the ground, cosmic-ray rejection is performed using contiguous science frames employing the method described by \citet{dokkum}. Figure~\ref{fig:fig1} presents the flowchart of the pipeline. 

CONTROL (available for download at {\it https://github.com/agsreejith/CONTROL}) is written in IDL 8.5.1 and is designed to work on both UNIX and WINDOWS systems. The pipeline has been tested for compatibility with previous IDL versions up to version 7.1. The intermediate and final output files generated by the pipeline are in FITS format, similar to all CUTE data products. CUTE data levels and file formats are described in detail in the next sections. Each raw file can be processed individually by the pipeline, but we recommend running the pipeline on one complete CUTE visit of a given target to generate light curves.

\subsection{Data levels}
According to their reduction step, CUTE data are sub-divided into five data levels.\\
\noindent Level 0: Data from spacecraft (including housekeeping).\\
Level 1: Raw FITS files.\\
Level 2: Intermediate FITS files (hot/bad pixel, bias, dark, and cosmic ray).\\
Level 3: Calibrated FITS files.\\
Level 4: Light curves, and other science products.

All 2D FITS files from level 2 and higher contain a single header and three extensions, namely data, their errors, and data quality. All 1D files are ASCII tables. Each extracted 1D spectrum file comprises data, data errors, data quality, background, background error, and background data quality. Each 1D background subtracted spectrum file comprises data, data error, and data quality. Each 1D wavelength calibrated spectrum file comprises wavelength, data, data error, data quality. Each 1D flux calibrated spectrum file comprises wavelength, flux, flux error, data quality. Bad-quality frames and/or pixels are flagged in the data quality arrays as a widget being 0 for good pixels and 1 for bad pixels. The light curves files are made of ASCII columns giving observing time at the beginning of each observation in seconds from the first observation of the sequence, while the subsequent columns give the flux integrated over pre-selected wavelength bins (see below) and their uncertainties.
\section{Architecture and Operation of CONTROL}\label{sec:architecture}
\subsection{Creating level 0 data on board}
CUTE carries out a set of on-board processing on the CCD frames obtained. These algorithms are part of the instrument Flight Software (FSW) written in C\texttt{++}. The FSW also uses on-board lookup tables to store the most up-to-date parameters/variables used by the software functions, which can be updated by a table load commanded from the ground via UHF. The individual algorithms also generate information turned into the meta data packets, which are finally placed into the FITS header on the ground. These include orbital and environmental parameters (temperature, attitude, etc.) as well as image statistics in the overscan regions of the CCD and regions of interest in the frames, along with information about the spectral position and width. 

The onboard image processing has three sets of algorithms that can be carried out on the CCD data to create 2D data products. The primary data product is the trimmed 2D (TRIM2D) data, which is created by trimming the full CCD image of $515\times2200$ pixels around the center of the spectrum, with an extra option to trim away the overscan columns. The center value was identified during payload commissioning and is stored as a table value. The height of the trimmed frame is also determined by the table value, which is optimized considering the spread of the spectrum in the cross-dispersion direction and the resulting data volume. We also have additional algorithms acting on the 2D frames intended for reducing the data volume for faster downlinks if needed. The {\it outspec2D} function combines three raw science frames by computing the median values across each pixel, and then extracts a box from the median output around the centeroid to create a trimmed median 2D frame. The {\it bin2D} algorithm reduces the data volume of individual raw frames by median combining every 4x4 pixels into a single value. 

The remaining set of on-board algorithms is intended to be used in case of failure of S-band communication. These set of algorithms carries out on-board data reduction, including the creation of reference bias and dark frames to be stored on board in the SDRAM, and the extraction to create 1D spectra to be transferred to the ground through the UHF antenna. The master bias frame is created on board by median combining a set of bias frames; this algorithm also estimates the readout noise. The master dark frame is computed after bias subtraction and medianing of single dark frames; this algorithm also generates an uncertainty dark frame to be used for error propagation in the spectral extraction. Both functions make use of the on-board bad pixel map to correct for the previously detected CCD bad pixels. Bad pixels are corrected by replacing them with the median value computed from the surrounding (3$\times$3 pixels) ``good'' pixels. 

The science frames are first bias- and dark-corrected, along with propagation of the relative error frames. Cosmic rays are then identified and corrected as follows. The algorithm defines pixels affected by cosmic rays as those for which $|S_{bd}(i)$ - median($S_{bd}(i)$)$| >$ 5$\times\sigma$($S_{bd}(i)$), where $S_{bd}(i)$ is the $i$th small region (4$\times$4 pixel) in the bias- and dark-subtracted science frame ($S_{bd}$) and $\sigma$($S_{bd}(i)$) is the standard deviation of the 16 pixels composing $S_{bd}(i)$. Once pixels affected by cosmic rays are identified, the algorithm corrects them in the same way as for the bad pixels correction. The spectra encoded in the cosmic ray corrected science frames are then extracted by co-adding a predefined number of rows with respect to the spectrum centroid in the cross dispersion direction. The cross dispersion profile is estimated dynamically on each bias- and dark-corrected frame by computing for each CCD row (i.e. along the dispersion direction) the median value. The distribution of the median values along the CCD column gives then the shape of the cross-dispersion profile and the peak of the distribution provides the location of the centroid. The noise associated to each frame is also propagated as part of the extraction procedure. As a result of these algorithms, the satellite sends to the ground the 1D extracted spectra, their uncertainties, and the cross-dispersion profile across the entire image.

\subsection{CONTROL: Creating level 1 data}  
The Level 1 data are created by the Software Data Systems team at LASP by decoding the telemetry packets from the S-band and/or the UHF antenna. Each data product that is downloaded as part of data transfer results in two sets of telemetry packets - one for the data product-specific metadata and one for the product itself. The CUTE command and telemetry database (CTDB) maps the various data products and their metadata to their respective Application Process Identifier (APID) in these packets. The CTDB also specifies the total length of each of these packets as well as their byte-wise content description. The APIDs, along with the frame and packet number fields in the packets, are used to group the individual data products. The decoded packets are then used to reconstruct the Level 1 data, which are in FITS format. The header fields are identified from the corresponding metadata packets and they include image statistics along with essential spacecraft data such as attitude, temperature, and observation time stamp. CONTROL takes over after the initial FITS files are created and processes them to create archival ready raw FITS files. This part of CONTROL takes care of trimming the overscan region, populating header keywords, and accounting for missing pixels in data. For bias and dark frames the missing pixels are filled by interpolating with the surrounding pixels. CONTROL also updates the data quality array to account for missing pixels (see below). 

\subsection{CONTROL: Converting level 1 to level 3 data.}\label{sec:control_gd}
The aim of CONTROL, which is publicly available code, is to create 1D wavelength and flux calibrated spectra together with the transit light curves, considering predefined wavelength regions of interest. The pipeline can be executed in the following three modes. In ``default'' mode, the pipeline requires spectrum width and background trace as inputs; for all other parameters the default values are used. In this mode, the pipeline also uses default calibration files stored in the calibration folder (available with the software distribution). In ``autorun'' mode, the pipeline calculates the spectrum width of extraction automatically (see section~\ref{sec:extr}) and the background region is assumed to be located 30 pixels away from the centroid of the spectrum on either sides. All other parameters are as in ``default'' mode. The preferred mode of operation is called ``expand'', where the pipeline operations are carried out by providing a parameter file (an example is provided with the software distribution), which governs the actions of the pipeline. The content of the parameter file is described in detail in Appendix \ref{sec:parfile}.  The code carries out internal checks on the input parameter file to spot user errors and it uses the default values wherever they occur (e.g. typos in the input parameter names). Also, the code returns an error if it does not find any required parameter value. The logfile gives users the option to double-check the input parameters and inspect the cause of the eventual errors.

The pipeline has a modular approach to ease modifications. The execution of the different modules of the pipeline is controlled by the {\it steps} keyword, with which the user can decide to run a single or multiple steps along the operational flow of the pipeline. The individual options available are discussed in detail in Appendix \ref{sec:parfile} together with the details of the input parameter file. Before going through the assigned tasks, the pipeline carries out a general check of the file locations provided in the parameter file. The pipeline performs an automatic file classification based on the {\it FILETYPE} keyword present in the FITS header of each file. The user can also exclude certain files from processing by including their name in the ``badfiles.txt'' file located in the data folder. The pipeline also produces an extensive log file to give users information about the performed steps, adopted and computed parameters, and other details of the pipeline execution, as well as an error log file describing the encountered errors. The individual modules of the pipeline are described below.

\subsubsection{Hot and Bad pixel correction}
This module removes hot and bad pixels from all frames using the provided input hot/bad pixel map (a FITS file with the same size as the input CCD frame with values of 0 for good pixels and 1 for bad pixels). The removal algorithm is based on either interpolation or average, as set by the user in the parameter file. The default option is interpolation, which is the one employed in case of no user input. Interpolation or averaging is carried out first over an entire row and then over an entire column. The process is carried out in both directions to avoid problems caused by hot/bad rows/columns. The location of hot/bad pixels is flagged in the data quality extension associated with each file.

\subsubsection{Bias module}
The bias module creates the master bias frame to employ during the reduction by combining individual bias frames and updates the data quality flag where necessary. The master bias is created as a mean, median, or mode of the single bias frames following what is given in the input configuration file with median being the default method. This module further enables the use of a rejection algorithm for both pixels and frames in order to remove outliers. The rejection algorithm looks for pixels deviating more than a user-defined threshold from the median bias value. If the number of pixels per frame identified by the rejection algorithm is more than 0.1\% of the total number of pixels, the corresponding frame is excluded from the generation of the master bias.

\subsubsection{Dark module}
The dark module creates the master dark frame to employ during the reduction by combining individual bias-subtracted dark frames and updates the data quality flags where necessary. The master dark is created as a mean, median, or mode of the single dark frames, depending on what given by the user in the input configuration file. Median combine is the default. This module further enables one to use a rejection algorithm identical to that present in the bias module.

\subsubsection{Flat module}
The flat module creates master flat frames by combining individual bias- and dark-subtracted flat frames and updates the data quality flags where necessary. The master flat is created a mean, median, or mode, depending on what given by the user in the input configuration file, of the single flat frames. Median combine is the default. Also this module enables the use of a rejection algorithm similar to that present in the bias module. The limitations of the scope of a CubeSat mission made a detailed in-band flat-field acquisition program infeasible on the ground and flat-fields cannot be obtained in orbit, given the lack of calibration lamp and suitable sky fields at UV wavelengths; therefore flat fielding of CUTE data is not currently implemented. For this reason, pixel-to-pixel variations are left to be partially accounted for by the flux calibration. Although this module is not used for the reduction of CUTE data, we made it an integral part of the pipeline and described it here so that CONTROL can be used to reduce data also from other, future missions carrying on-board a payload not too dissimilar from that of CUTE (i.e. slit spectrograph with CCD detector).

\subsubsection{Spectrum Initialization }
This module carries out bias subtraction, dark correction, and flat fielding of the science frames employing the master bias, master dark, and master flat, respectively. Users can also provide the master calibration files of their choice through the parameter file. In the absence of a master bias, master dark, and/or master flat, the pipeline assumes the bias level and/or, dark level, and/or flat field level to be equal to zero, zero, and one, respectively. As part of this module, the pipeline also identifies the location of saturated pixels and updates the relevant data quality pixels.

\subsubsection{Cosmic ray correction}
This module removes cosmic rays using the LAcosmic algorithm \citep{dokkum}, which is based on a variation of the Laplacian edge detection capable of identifying cosmic-rays of arbitrary shape and size by the edge sharpness. The user has the option to carry out cosmic ray rejection before or after dark correction. The default option of cosmic ray correction is after dark subtraction. Most of the keywords required for LAcosmic are set by the pipeline internally, but users have the option to set in the parameter file the {\it sigclip} keyword, which determines the level of cosmic ray clipping (threshold). Initial CUTE observations indicate that the optical {\it sigclip} value to use may differ from target to target.

\subsubsection{Spectrum identification, trace and extraction}\label{sec:extr}
The CUTE spectrograph disperses light in wavelength along the long axis of the CCD (i.e., the x-axis). The observations are carried out in such a way to keep the target star at the slit center, but the location of the spectrum along the y-axis of the CCD can be changed if needed.

This module of CONTROL identifies the location of the spectrum along the CCD y-axis (i.e., centroid), defines the trace region, and extracts the spectrum. The CUTE spectral trace is angled with respect to the CCD \citep{Egan2022}. Therefore, the position of the centroid as a function of pixel in the dispersion direction can be defined as
\begin{equation}
    {\rm centroid} = {\rm slope} \times\ {\rm pixel} + {\rm cent\_value}\,,
\end{equation}
where cent\_value is the centroid value corresponding to the 0th pixel along the CCD x-axis (on the left-hand side of the detector). The slope and cent\_value parameters are user inputs and they are regularly checked and updated during science operations. CUTE's   spectral trace slope remained constant so far since payload commissioning. The pipeline also allows the user to change the functional shape of the spectral trace in the dispersion direction to a higher-order polynomial, in which case the user has to enter the degree and coefficients of the polynomial. Instead, if the slope and cent\_value parameters are not given by the user, the pipeline computes them as described below. First the pipeline gets a first estimate of the position of the centroid by summing the flux across columns, binning at every 8 pixels in the dispersion direction. The rough centroid position is then assumed to be that of the pixel with the maximum count along the y-axis of the detector. This rough centroid information is then fitted with a polynomial of degree specified by the user (the default is one) to generate a pixel to centroid relation. The pipeline finally computes the centroid position for each pixel along the x-axis based on the coefficients of the polynomial fit. This procedure assumes that the target of observation is the brightest source in the spectrum.  CUTE targets were selected to avoid bright stars within 2' to avoid contamination. If this is not possible, the slit can be oriented by changing the roll angle, to remove brights stars from the slit field of view \citep{fleming}.

The spectral extraction is performed by collapsing the signal in the cross-dispersion direction, taking into account the spectrum's width and shape in the cross dispersion direction. The cross dispersion angle has negligible effects on the resolution and thus it is assumed to be aligned along the CCD columns. The process of defining the spectral extraction region depends on the extraction method given by the user in the input parameter file. These are the options for the spectral extraction and each option also takes care of propagating the uncertainties accordingly. 
\begin{itemize}
    \item {\bf Simple} $-$ In this case, the extraction region is a box with a fixed width and slope. This is the default option of the pipeline and the user needs to provide the width and slope in the input parameter file. The pipeline extracts the 1D spectrum by summing up rows within the defined extraction region.  
    \item  {\bf Fixed} $-$ This method is similar to the previous one, but it enables one to set different values for the upper and lower width of the extraction box from the centroid line. The user needs to provide both width values and the slope in the input parameter file. The extraction of the 1D spectrum is then identical to that of the previous case.
    
    \item {\bf Variable} $-$ In this case, the user defines the extraction width column-by-column by setting a threshold value. The lower and upper extraction limits are different and define the location in each column in which the pixel counts are above the threshold given by the user times the maximum pixel count in that column. Possible outliers lying outside of the desired extraction box are identified and removed through a smoothing function. The pipeline then fits a second degree polynomial to the location of the upper and lower limits of each extraction window along the x-axis of the CCD to obtain a smooth extraction region. The 1D spectrum is then obtained by summing up rows in the so-defined extraction region.
    
    \item {\bf Function} $-$ This extraction method fits a function in the cross-dispersion direction, considering a region of about 40 pixels around the centroid value and defines the extraction boundaries as those having a pixel value higher than a fraction (given by the user) of the peak value. In practice, this method differs from the previous one only in the definition of the peak value. The available fitting functions are Gaussian, Lorentzian, and Moffat. The 1D spectrum is then obtained as in the previous cases.
\end{itemize}

\begin{figure}[ht!]
\begin{center}
\includegraphics[width=0.99\textwidth,height=\textheight,keepaspectratio]{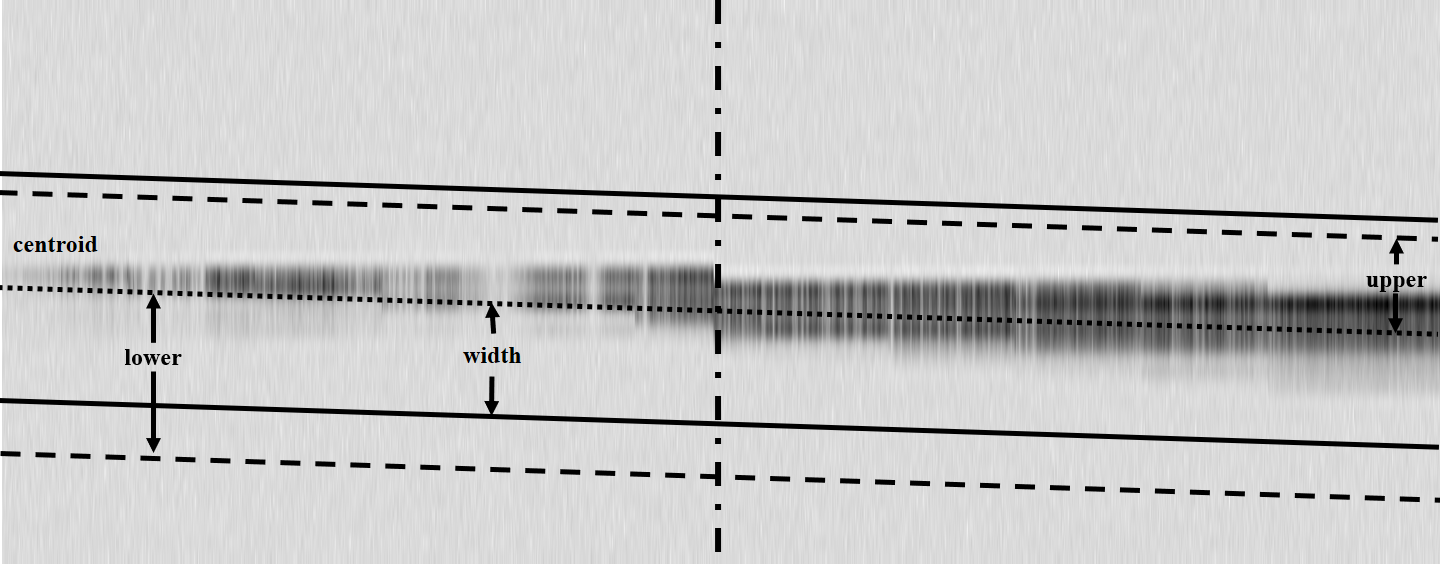}
\vspace{0.5cm}
\includegraphics[width=0.99\textwidth,height=\textheight,keepaspectratio]{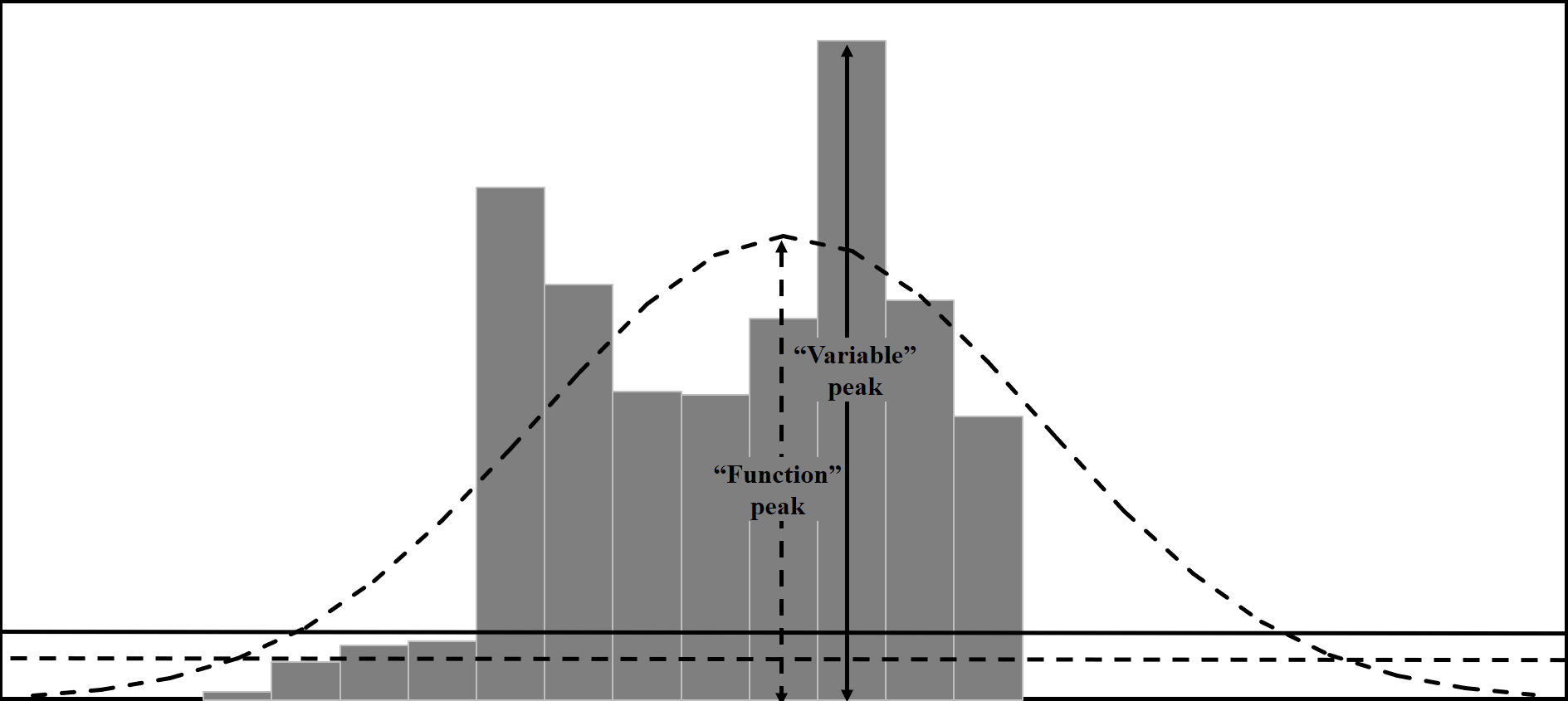}
\caption{Schematic representation of the different trace and extraction methods available in CONTROL. Top: Portion of a simulated CUTE spectral image. The dotted line represents the centroid of the spectrum. The solid lines represent the extraction region as defined by the ``simple'' method, hence with a constant width (10 pixels) in both directions. The dashed lines represent the extraction region of the ``fixed'' method with different widths for the lower (15 pixels) and upper (8 pixels) trace boundaries. The vertical dash-dotted line indicates an example column along which the pipeline analyses the spectrum in the case of the ``variable'' and ``function'' extraction methods described in the bottom panel. Bottom: The shaded distribution presents the spectral counts in the cross-dispersion direction along the dash-dotted line in the top panel. The vertical and horizontal solid lines indicate the peak and extraction region in case the ``variable'' extraction method is selected with a threshold of 10\%. The vertical and horizontal dashed lines indicate the peak and extraction region in case the ``function'' extraction method is selected with a threshold of 10\%.}
\label{fig:fig2}
\end{center}
\end{figure}

To measure the background level simultaneously, this module extracts a spectrum in a region of the CCD that is specified by the user and that is not illuminated by the target or background stars. The pipeline enables one to use three methods to define the background extraction window. If the user provides one value (positive or negative) in the relevant field of the input parameter file, the pipeline sets a background extraction box identical to that of the target, but with its central position displaced above or below the centroid of the stellar spectrum by the amount (in pixels) provided by the user. If the user provides more than one number (positive or negative) then the pipeline sets multiple background extraction boxes (as many as the values in input) displaced from the centroid of the stellar spectrum by as much as given in the parameter input file. Furthermore, the user has the capability to set the width of each background extraction box. In case of crowded fields, the user has also the possibility to collapse the background extraction boxes to single rows identical to the target spectral trace. The pipeline then scales the widths of the background extraction boxes to that of the target to avoid over- or under-subtraction of the background. During operations, the position of the background extraction window varies from target to target and it is decided upon inspection and testing on simulated CUTE spectral images produced with the CUTE data simulator \citep{cutedrndl} that accounts for background stars on the basis of information extracted from the GAIA catalog \citep{gaiadr2}. This module also generates quick look images of the trace region. 

\subsubsection{Background subtraction}
This module subtracts the background from the extraction spectrum. Uncertainties and data quality flags are also propagated. In case the S-band antenna fails and 1D spectra are produced on board without background subtraction, then this becomes the first step carried out by the ground-based data reduction pipeline where it packages the 1D spectra without background subtraction for further steps.

\begin{figure}[h!]
\begin{center}
\includegraphics[width=\textwidth]{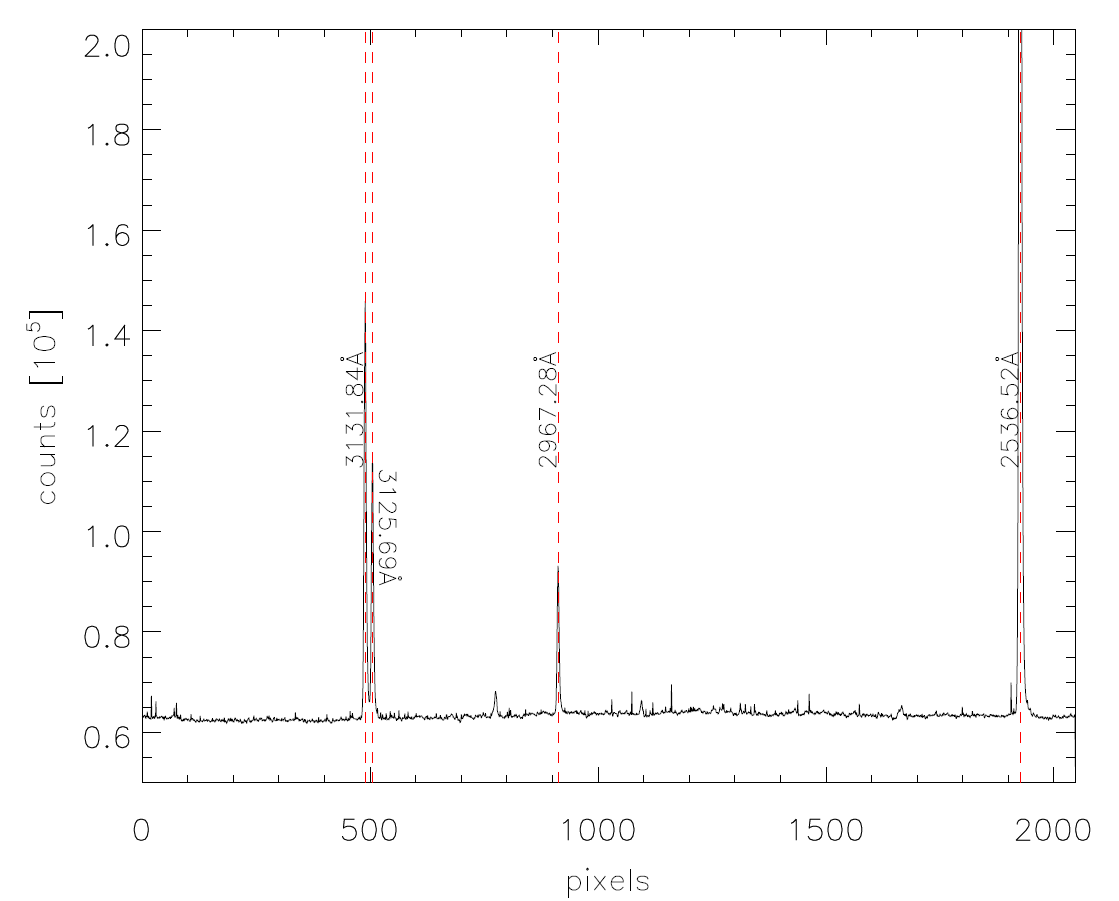}
\caption{Mercury pen-ray lamp spectrum obtained with CUTE during ground calibration.}
\label{fig:lamp}
\end{center}
\end{figure}

\subsubsection{Wavelength calibration}\label{sec:wavelength_calibration}

Wavelength calibration frames have been obtained on the ground during testing and calibration, and by default the pipeline employs the pixel to wavelength map obtained on the ground to perform wavelength calibration. However, during commissioning we obtained calibration targets spectra, which we use for wavelength calibration. For lamp spectra, such as that shown in Figure~\ref{fig:lamp}, the pipeline executes all steps untill background subtraction to obtain the calibrated two-dimensional image and one-dimensional lamp spectra. The pipeline considers first a basic wavelength solution of the form 
 \begin{equation}
 Wavelength=W_0+a_1 \times pixel + a_2 \times pixel^2  
 \end{equation}
where $W_0$ is the wavelength reference point and $a_1$ and $a_2$ are the quadratic fit coefficients to CUTE's wavelength solution. This first wavelength solution is needed to enable the pipeline to identify the lines present in the wavelength calibration frame. The pipeline then looks for emission lines in the calibration frame and fits Gaussians onto them to find their central locations, which are then converted into wavelengths on the basis of an input file containing the wavelengths of known emission features. The wavelengths corresponding to the center of emission lines are fitted to their pixel locations with a fourth degree polynomial to obtain a pixel-to-wavelength solution. This polynomial solution is used to generate a one-dimensional pixel to wavelength map. 

The pipeline has also the capability of generating a two-dimensional wavelength solution map for the entire detector. This is achieved by following the procedure described above in each row in the calibrated two-dimensional spectra. These updated wavelength solutions are stored in the output directory. However this implementation is not used in the specific case of CUTE. 

However, the pipeline enables the user to also employ a further calibration step to correct for rigid wavelength shifts. This additional step cross correlates the observed spectrum with a photospheric synthetic spectrum of the target star, given as input. The pipeline then finds the peak of the cross correlation function and uses it to correct for any rigid shift. This algorithm is similar to that employed by the IUE data reduction pipeline. Users also have the option to carry out auto correlation with the first observation in the set, instead of cross correlation with a synthetic spectrum.  We remark that cross correlation techniques have been shown to be applicable also to spectra at the resolution of the CUTE spectrograph \citep[e.g.,][]{bagnulo}

As part of the instrument monitoring operations, CUTE has so far observed three bright stars (HD 60178J, HD\,66811, and HD\,186882) to set the flux calibration, to correct for possible rigid wavelength shifts, and to identify the possible presence of wavelength-dependent deviations from the pixel-to-wavelength solution. Given the almost Sun-synchronous orbit of CUTE and to enable one carrying out these checks throughout the year, we have further selected bright, single, slowly rotating early G- to late F-type stars that are well spread in right ascension and have a declination between $-$60 and $+$60 degrees. The characteristics of these stars (i.e. bright, single, slowly rotating, early G- to late F-type) ensure that each spectrum is of high quality, contains a large number of lines, line broadening is dominated by instrumental broadening, and the position of the spectral lines is stable in time at the spectral resolution of CUTE. Table~\ref{table1} lists the stars that we identified for this task. To check for the stability of the wavelength solution, we also employ cross-correlation with previous UV observations or synthetic spectra.
\begin{table}
\begin{tabularx}{\textwidth}{X | c | c | c | c}
Identifier  	&RA(J2000) & Dec (J2000)& V mag & Spectral type   \\
\hline
HD 3229    	&	00 35 32.83 	&	-00 30 20.19	&   5.94	&	F5V	\\
HD 4813 	&	 00 50 07.59	&	 -10 38 39.58   &	5.19	&	F7V	\\
HD 8671    	&	 01 26 18.68	&	 +43 27 27.87   &	5.98	&	F7V	\\
HD 15335   	&	 02 28 48.49	&	 +29 55 54.33   &	5.90	&	 G0V	\\
HD 17051   	&	02 42 33.46 	&	-50 48 01.05	&   5.40 	&	F8V	\\
HD 22484    &	 03 36 52.14	&	 +00 23 58.54   &	4.30	&	F9IV-V	\\
HR 1538    	&	 04 48 32.53	&	 -16 19 46.15   &	5.75	&	F8V	\\
HD 34721   	&	 05 18 50.47	&	 -18 07 48.18   &	5.96	&	G0V	\\
HD 45067   	&	 06 25 16.55	&	 -00 56 45.18   &	5.90	&	F9V	\\
HD 50692   	&	 06 55 18.67	&	 +25 22 32.50   &	5.75	&	G0V	\\
HD 58855   	&	 07 29 55.96	&	 +49 40 20.86   &	5.36	&	 F6V	\\
HD 60532   	&	 07 34 03.18 	&	-22 17 45.84	&   4.39	&	F6IV-V	\\
HD 67228 	&	 08 07 45.85	&	 +21 34 54.53   &	5.93	&	G1IVb	\\
HD 82328   	&	 09 32 51.43 	&	 +51 40 38.28 	&   3.18	&	F7V	\\
HD 84117   	&	 09 42 14.42	&	 -23 54 56.05   &	4.94	&	F9V	\\
HD 84737   	&	 09 48 35.37	&	 +46 01 15.63   &	5.10	&	G0.5Va 	\\
HD 102870  	&	 11 50 41.72 	&	 +01 45 52.99	&   3.60	&	F9V	\\
HD 114710  	&	 13 11 52.39 	&	 +27 52 41.45	&   4.25	&	F9.5V	\\
HD 141004 	&	 15 46 26.61	&	 +07 21 11.04	&   4.42	&	G0-V	\\
HD 143790  	&	16 03 34.36 	&	-32 00 01.92	&   5.99	&	F5IV/V	\\
HD 156098  	&	17 17 03.64 	&	-32 39 46.22	&   5.53	&	F6V	\\
HD 157214  	&	17 20 39.57 	&	 +32 28 03.87   &	5.39	&	G0V	\\
HD 169830  	&	18 27 49.48 	&	-29 49 00.70	&   5.90	&	F7V	\\
HD 198390  	&	20 49 37.77 	&	+12 32 42.45	&   5.99	&	F5V	\\
HD 222368  	&	 23 39 57.04 	&	 +05 37 34.65	&   4.12	&	F7V	\\
HD 60178J   & 07 34 35.87       & +31 53 17.82      & 1.58      & A1V\\
HD66811     & 08 03 35.05       & -40 00 11.33      & 2.25      & O4\\
HD186882    & 19 44 58.48       & +45 07 50.92      & 2.87      & A0IV\\

\hline
\end{tabularx}
\caption{\label{table1}Potential CUTE wavelength calibration targets.}
\end{table}

\subsubsection{Flux calibration}
This module converts the standard CCD counts \,s$^{-1}$\,pixel$^{-1}$ into the flux unit erg\,cm$^{-2}$\,s$^{-1}$\,\AA$^{-1}$. This module takes as input a file containing the instrument response function (sensitivity function)
obtained following the observation of flux calibration standard stars. For this task, we employ the standard stars listed in Table~\ref{table2} that have been drawn from the ESO spectrophotometric standards list and for which IUE and HST NUV observations are available. The selection of this sample of stars was based on a magnitude cut off at $V$\,=\,11\,mag and on their location in the sky, for which we followed considerations similar to those described in Section~\ref{sec:wavelength_calibration}. Most of the previous UV missions have cut off around 3000\AA, while ground-based observations reach wavelengths as short as 3500\AA. This makes flux calibration in the 3100--3400\,\AA\ range difficult, which is why the CUTE flux calibration at long wavelengths is subject to some uncertainty.

From the observations of the flux calibration standard stars, we generate the CUTE instrument response as a function of wavelength. Therefore, we obtain the CUTE instrument response function by dividing the CCD counts in the extracted and wavelength-calibrated 1D spectrum by the NUV flux-calibrated spectrum of the standard calibration source obtained with a different instrument (e.g., IUE, HST) convolved to CUTE resolution.  The response function is further interpolated and smoothed to avoid low-flux effects in the core of stellar spectral lines. For CUTE, since no ground-based flat-field has been obtained and it is not possible to construct a flat-field image in space, the flux calibration procedure corrects for CCD pixel-to-pixel variations.

\begin{center}
\begin{table}
\begin{tabularx}{\textwidth}{X | c | c | c | c}
Identifier & RA (J2000) & Dec (J2000) & V mag & Spectral type \\
\hline
HR153       & 00 36 58.28 & +53 53 48.87 & 3.66  & B2IV \\
HR718       & 02 28 09.56 & +08 27 36.22 & 4.28  & B9III \\
HD60753     & 07 33 27.32 & -50 35 03.31 & 6.70  & B3IV\\
HD93521     & 10 48 23.51 & +37 34 13.09 & 7.04  & O9Vp\\
HR4468      & 11 36 40.91 & -09 48 08.09 & 4.70  & B9.5V\\
HR4554      & 11 53 49.85 & +53 41 41.13 & 2.44  & A0V\\
Feige66     & 12 37 23.52 & +25 03 59.87 & 10.50 & sdO\\
HR5191      & 13 47 32.44 & +49 18 47.76 & 1.86  & B3V\\
BD+33d2642  & 15 51 59.86 & +32 56 54.8  & 10.81 & B2IV\\
HR7001      & 18 36 56.34 & +38 47 01.28 & -0.01 & A0V\\
HR7950      & 20 47 40.55 & -09 29 44.78 & 3.78  & A1V\\
BD+28d4211  & 21 51 11.07 & +28 51 51.8  & 10.51 & O\\
BD+25d4655  & 21 59 42.02 & +26 25 58.1  & 9.76  & O\\
HR8634      & 22 41 27.72 & +10 49 52.91 & 3.40  & B8V\\
HD 60178J   & 07 34 35.87 & +31 53 17.82 & 1.58  & A1V\\
HD66811     & 08 03 35.05 & -40 00 11.33 & 2.25  & O4\\
HD186882    & 19 44 58.48 & +45 07 50.92 & 2.87  & A0IV\\
\hline
\end{tabularx}
\caption{\label{table2}Potential CUTE flux calibration targets.}
\end{table}
\end{center}

\begin{figure}[h!]
\begin{center}
\includegraphics[width=\textwidth]{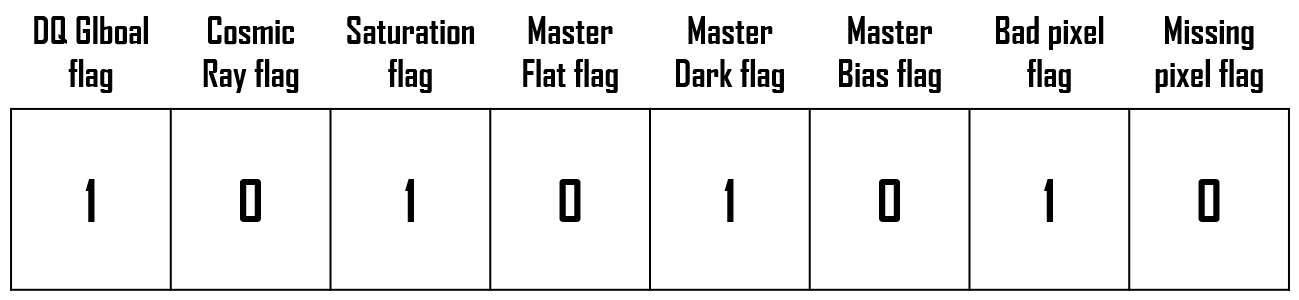}
\caption{An example of data quality byte for a particular pixel. The bit corresponding to the data quality flag associated with each data reduction procedure is labeled. This example is for a pixel with the following issues: it is a flagged hot/bad pixel, there are data quality issue in master dark frame and is also a saturated pixel.}
\label{fig:fig3}
\end{center}
\end{figure}

\begin{figure}[h!]
\begin{center}
\includegraphics[width=\textwidth]{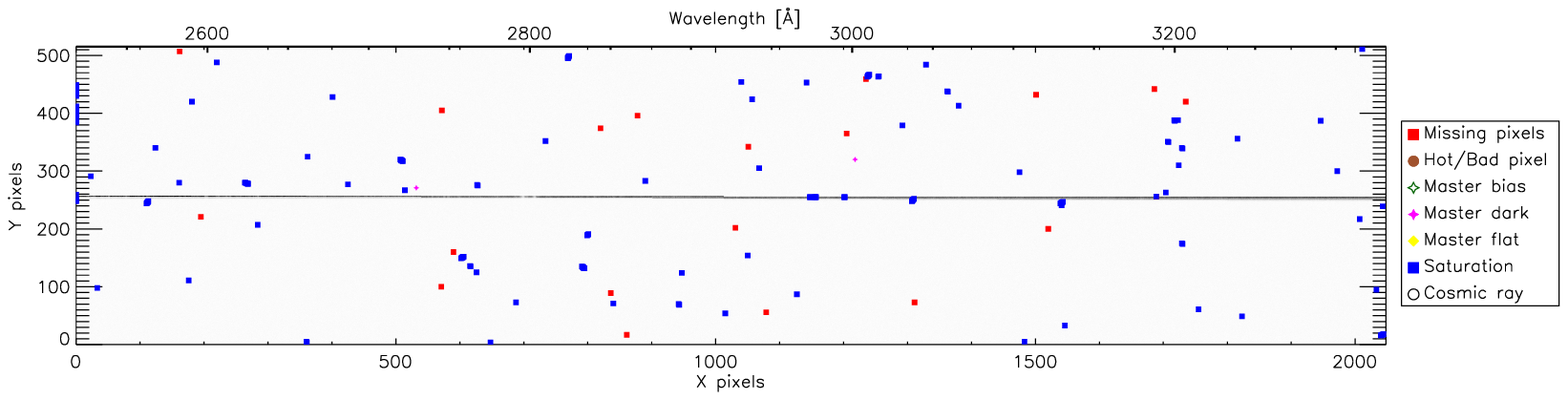}
\caption{An example of data quality image generated by the pipeline.}
\label{fig:fig4}
\end{center}
\end{figure}

\subsection{Predefined level 4 data: light curves}
This module generates light curves from a particular CUTE visit integrating over pre-defined wavelength regions in the CUTE band. By default, this module produces light curves integrating over seven wavelength regions: one integrating across the full CUTE wavelength band (i.e., white light curve), three light curves integrating across three equally wide regions across the whole CUTE band, and three light curves integrating across regions covered by features of interest for the study of exoplanetary upper atmospheres \citep[see][]{fossati2018b,fossati2020}, namely Mg{\sc i} (2850$-$2854\,\AA), Mg{\sc ii} (2793$-$2805\,\AA), and Fe{\sc ii} (2583$-$2587\,\AA) \citep{cubillos2020}. The pipeline stores all light curves in a single file with the corresponding time information (time in seconds from the mid-time exposure of the first observation) in the first column. 

\subsection{Error propagation}
Each step of the pipeline propagates the uncertainties and updates the data quality flags. Here we describe the data reduction steps mathematically, focusing in particular on the error propagation. 

Let $B$, $D$, $F$, and $I$ be the bias, dark, flat, and raw science frames respectively, and $\sigma B$, $\sigma D$, $\sigma F$, and $\sigma I$ their uncertainties. To keep the equations contained, we do not write the processes the pipeline employs to generate master calibration files and therefore assume that we have just one bias, one dark, and one flat frame. The bias-subtracted dark, flat, and science frames are
 \begin{equation}
 D_b = D - B
 \end{equation} 
 and
  \begin{equation}
 F_b = F - B\,,
 \end{equation} 
and
 \begin{equation}
 I_b = I - B\,,
 \end{equation} 
hence their uncertainties are
 \begin{equation}
 \sigma D_b=\sqrt{{D_{be}}+R^2}
 \end{equation}
 and
  \begin{equation}
 \sigma F_b=\sqrt{{F_{be}}+R^2}\,,
 \end{equation}
 and
 \begin{equation}
 \sigma I_{b}=\sqrt{I_{be}+R^2}\,.
 \end{equation} 
In the above Equations,
\begin{equation*}
D_{be} = \begin{cases}
\frac{D_b}{G} & \quad \frac{D_b}{G} \geq 0 \\
0 & \quad \frac{D_b}{G} < 0
\end{cases}
\end{equation*}
and
\begin{equation*}
F_{be} = \begin{cases}
\frac{F_b}{G} & \quad \frac{F_b}{G} \geq 0\,, \\
0 & \quad \frac{F_b}{G} < 0\,,
\end{cases}
\end{equation*}
and
\begin{equation*}
I_{be} = \begin{cases}
\frac{I_b}{G} & \quad \frac{I_b}{G} \geq 0\,, \\
0 & \quad \frac{I_b}{G} < 0\,,
\end{cases}
\end{equation*}
where $R$ is the CCD readout noise and $G$ is the CCD gain, which is defined as the ratio of CCD counts (ADUs) to photoelectrons (e.g., if the gain is equal to 2.1, for every 3 photoelectrons one would record 6.3 ADUs). The division by the gain ensures that the CCD counts are converted into photons, which then enables one to apply photon noise statistics in the error propagation. Following the same notation, the bias and dark subtracted flat and science frames, and their uncertainties, are
  \begin{equation}
 F_{\rm bd} = F_{\rm b} - D_{\rm b}
 \end{equation} 
 \begin{equation}
 I_{\rm bd} = I_{\rm b} - D_{\rm b}
 \end{equation} 
and
 \begin{equation}
 \sigma F_{\rm bd} = \sqrt{\sigma F_{\rm b}^2 + \sigma D_{\rm b}^2}
 \end{equation} 
 \begin{equation}
 \sigma I_{\rm bd} = \sqrt{\sigma I_{\rm b}^2 + \sigma D_{\rm b}^2}\,.
 \end{equation} 
The normalised flat field and its uncertainty are
 \begin{equation}
 F_{\rm bdn} = \frac{F_{\rm bd}}{n}
 \end{equation} 
 and
 \begin{equation}
 \sigma F_{\rm bdn} = \frac{\sigma F_{\rm bd}}{n}\,,
 \end{equation} 
which are then used to obtained the fully calibrated 2D science frames
 \begin{equation}
 I_{\rm bdf} = \frac{I_{\rm bd}}{F_{\rm bdn}}
 \end{equation} 
and their uncertainties
 \begin{equation}
 \sigma I_{\rm bdf} = \sqrt{\left(\frac{\sigma I_{\rm bd}}{F_{\rm bdn}}\right)^2 +\left(\frac{I_{\rm bd}}{F_{\rm bdn}^2}\cdot\sigma F_{\rm bdn}\right)^2}\,.
 \end{equation}

Finally, for each wavelength bin, the uncertainties of the 1D extracted spectrum ($\sigma_{1D}$) are given by
 \begin{equation}
 \sigma_{\rm 1D} = \sqrt{\sigma_{\rm ext}^2 + \sigma_{\rm bkg}^2}\,,
 \end{equation}
where $\sigma_{\rm ext}^2$ is the uncertainty associated to the extracted spectrum and $\sigma_{\rm bkg}^2$ is the uncertainty associated to the background spectrum, which are computed as 
 \begin{equation}\label{eq:error1}
 \sigma_{\rm ext} = \sqrt{\sum\limits_{\rm bottom}^{\rm top}\sigma I_{\rm bdf}^2}
 \end{equation}
and
  \begin{equation}\label{eq:error2}
 \sigma_{\rm bkg} =  \frac{\sqrt{\sum\limits_{n}^{}\sum\limits_{\rm bkg\_bottom_n}^{\rm bkg\_top_n}\sigma I_{\rm bdf}^2}}{\frac{Npix_{\rm ext}}{Npix_{\rm bkg}}}\,.
 \end{equation}
In Equation~\ref{eq:error1}, the sum is over all pixels along a column from the bottom to the top of the extraction window. In Equation~\ref{eq:error2}, the first sum is over the $n$ background extraction windows, while the second sum is over all pixels along a column from the bottom to the top of each background extraction region. The denominator in Equation~\ref{eq:error2} contains the number of rows summed over the spectrum extraction window ($Npix_{\rm ext}$) and the number of rows summed over all background extraction windows ($Npix_{\rm bkg}$), and it accounts for the possible different sizes of the target and background extraction windows.

\begin{figure}[b!]
\begin{center}
\gridline{\fig{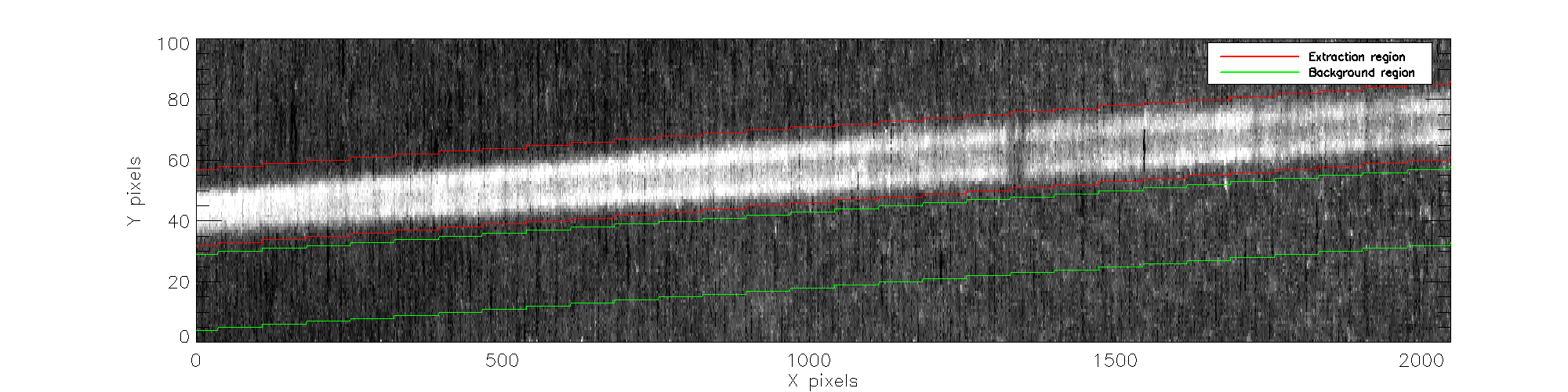}{0.7\textheight}{(a)}}
\gridline{\fig{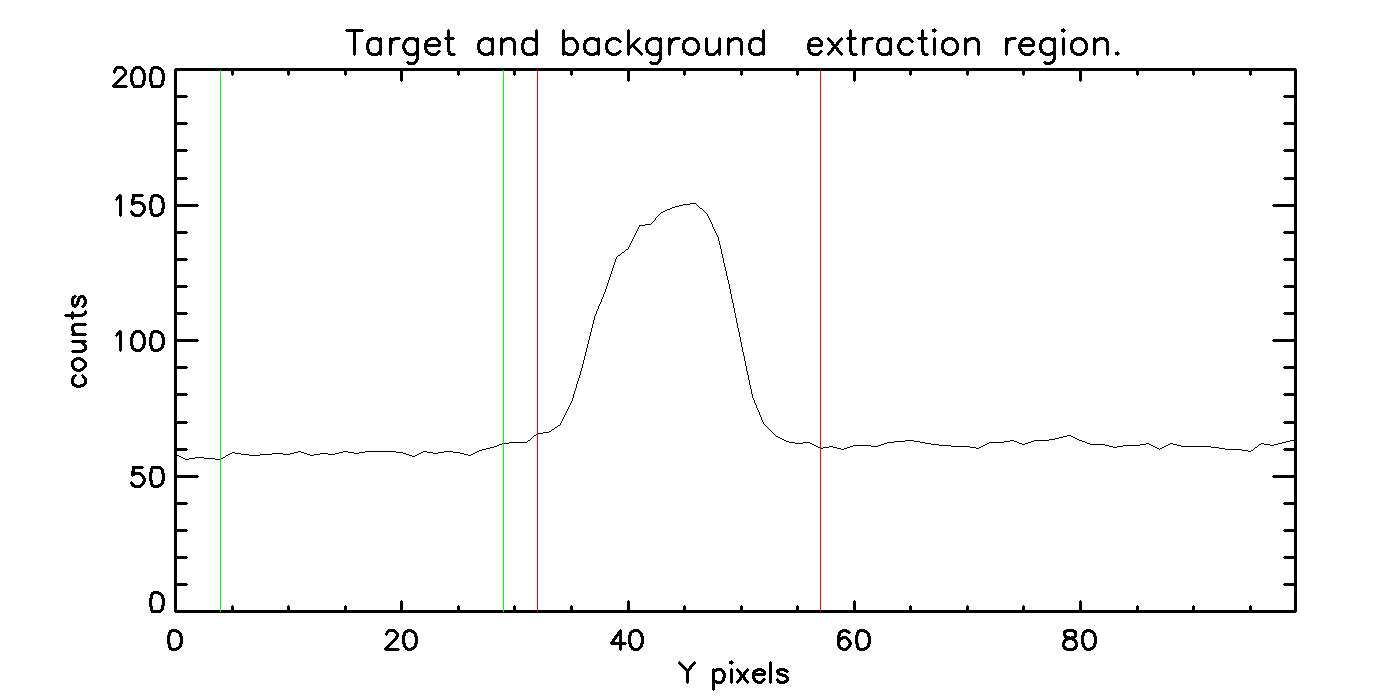}{0.49\textwidth}{(b)}
          \fig{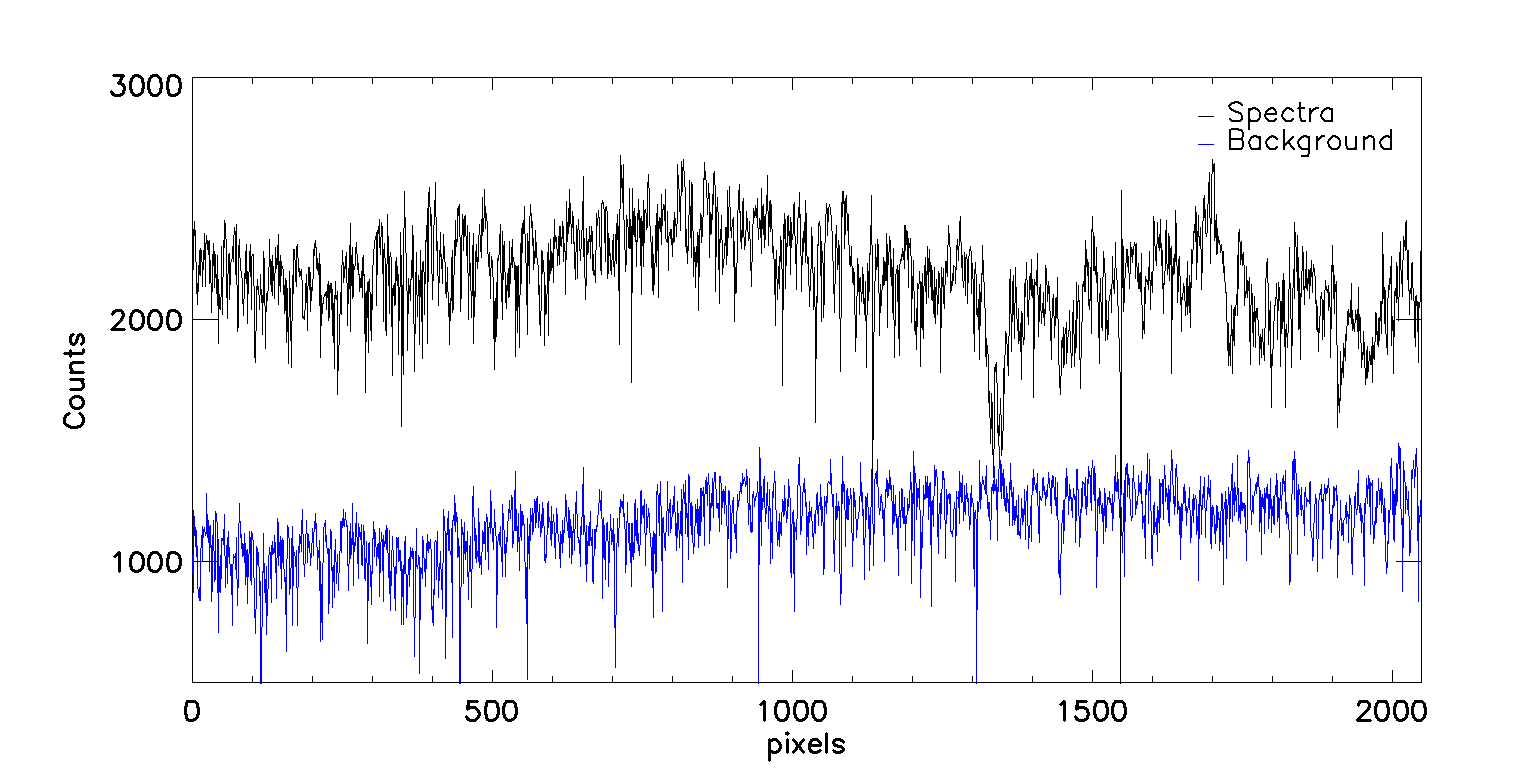}{0.49\textwidth}{(c)}}
\gridline{\fig{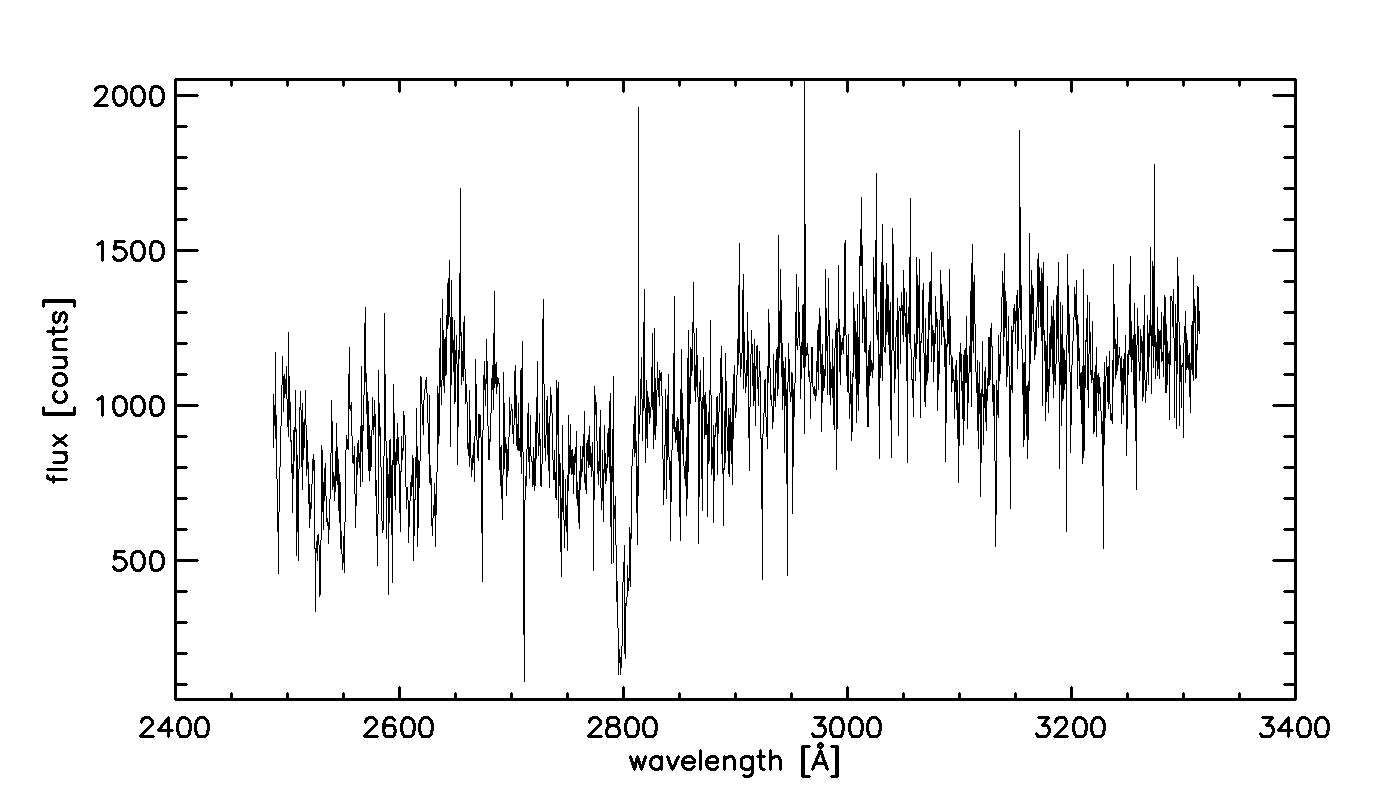}{0.49\textwidth}{(d)}
          \fig{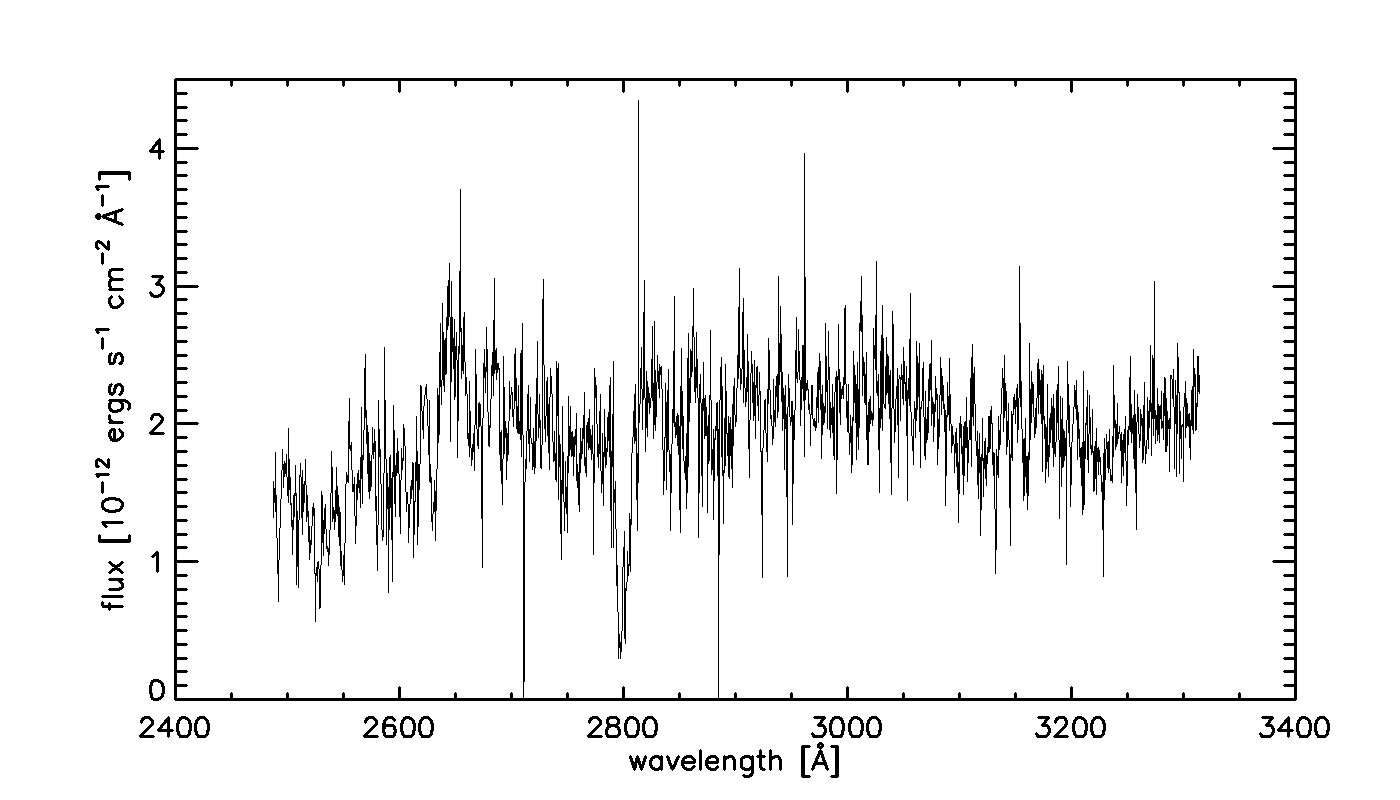}{0.49\textwidth}{(e)}
         }
\caption{Sample of output figures produced by the pipeline on the basis of CUTE observations of the hot Jupiter WASP-189b, {\bf (a):} CUTE spectral image (bias and cosmic ray corrected) showing the extraction region that is contained between the two red lines. The background region is shown by green lines. {\bf (b):} Median CCD counts in the cross-dispersion direction for 100 pixels. The red and green vertical lines indicate the maximum extend of target and background extraction windows, respectively. {\bf (c):} Extracted spectra and background. {\bf (d):} CUTE wavelength calibrated spectrum of a single 300 seconds exposure. {\bf (e):} CUTE wavelength and flux calibrated spectrum of a 300 seconds exposure.}
\label{fig:fig5}
\end{center}
\end{figure}

\begin{figure}[h!]
\begin{center}
\includegraphics[width=\textwidth]{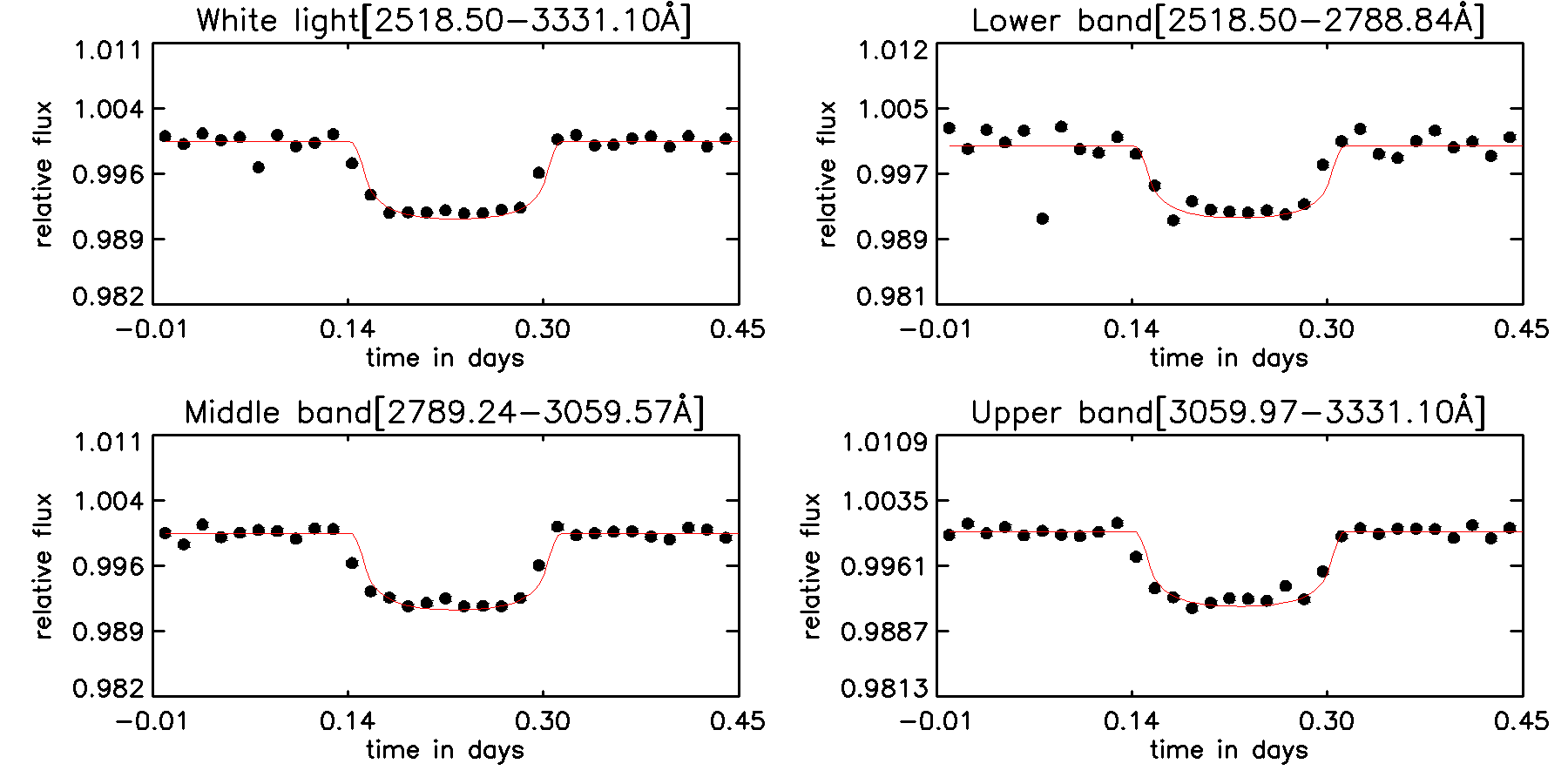}
\caption{ Sample light curves produced by the pipeline on the basis of simulated CUTE observations of the hot Jupiter KELT-9b, assuming a constant planetary radius with wavelength equal to that measured in the optical. The red curve shows the input light curve to the simulation. Each light curve data point has been rebinned four times. Light curves from actual CUTE observations will be presented in a future publication.}
\label{fig:fig6}
\end{center}
\end{figure}

\begin{figure}[h!]
\begin{center}
\includegraphics[width=\textwidth]{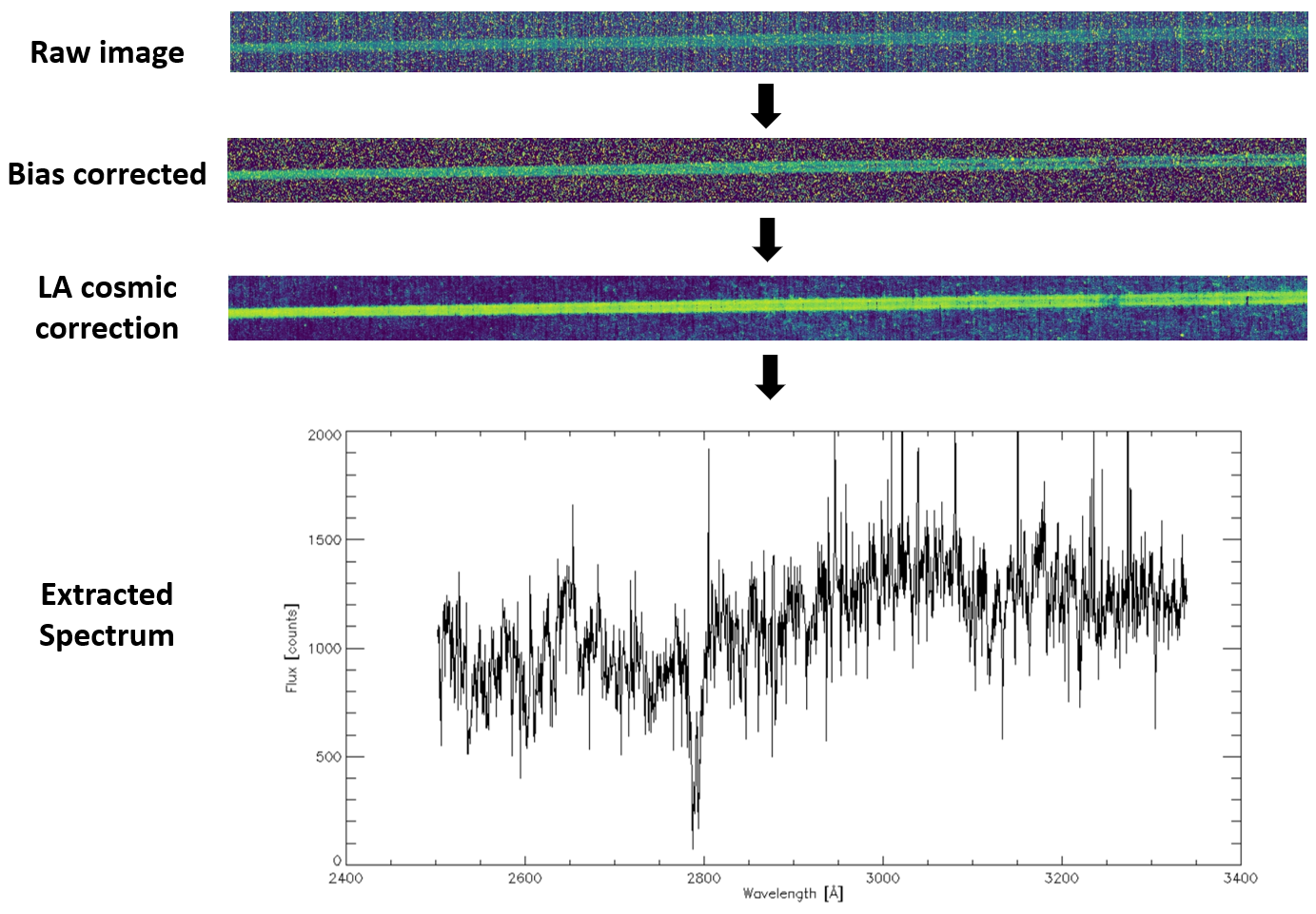}
\caption{Example data reduction steps carried out by the pipeline on flight data.}
\label{fig:all}
\end{center}
\end{figure} 
 
\subsection{Data quality propagation}

All the locations of hot and bad pixels even after correction are flagged as bad data pixels. Bias, dark and flat files have similar data quality checks. For each of these master files created, pixels that deviate from the median value by a certain threshold are flagged. We are currently using a 5$\sigma$ threshold as the default value, but this is a quantity that the user can control through the parameter file. Saturated pixels are also flagged as bad pixels. The saturation limit of the CCD can be set by the user through the parameter file. For science files, locations of cosmic ray hits are flagged as bad data. The data quality steps inherited from the previous data reduction/processing steps are also conserved. We plan to add data quality flags for CCD defects and pixels affected by scattered light. NaN data values in pixels are also accounted for after each step in the pipeline.

\subsection{Additional files}
The pipeline by default produces the run and error logs for each run. These files are created following \citet{perley} and are additional to what the pipeline displays on terminal. The run log file is located in the data folder and it contains a description of each step carried out by the pipeline together with all the employed parameters, including the default parameters. The error log file lists the errors has encountered by the pipeline and gives users the possibility to check where an error occurred, acting as a troubleshooting guide. Together with the two log files, the pipeline also produces plots presenting the area considered to trace the spectrum, the extraction boxes (target and background regions), wavelength and flux calibrated spectra, and the light curves. The pipeline also generates a plot presenting how the target flux and its uncertainty changes with the size of the extraction window, enabling the user to set the most suitable extraction window that maximises the target flux, yet minimising the noise. This plot also helps fine-tuning the on-board data reduction procedure, if required. Figure~\ref{fig:fig5} and Figure~\ref{fig:fig6} present examples of the plots produced by the pipeline.

\subsubsection{Jitter files}
The pipeline also creates a housekeeping file for each observation containing the jitter information that is obtained from the attitude control system of the spacecraft with a cadence of 10 seconds. CONTROL reads this file to create a FITS table summarising the following information: time in Julian date, payload latitude and longitude, right ascension and declination of the pointing, position angle, and 3-axis jitter. The pipeline also updates the image header with jitter keywords and creates plots presenting how temperature, latitude, longitude, altitude, RA, Dec, Roll angle, and corresponding jitter vary with time during each observation. Missing data values due to telemetry issues are either filled in from image metadata, when available, or else assumed to be zero.

\section{Dealing with flight data}

The CONTROL pipeline has undergone rigorous module and end-to-end tests based on simulated data. However, the first in-flight data demonstrated that some modifications were necessary. Before and after each visit, we collect a set of bias frames that we use for bias correction. As a result of the loss of active cooling \citep[see][]{kevin2022,Egan2022}, hot pixels depend strongly on the detector temperature. Therefore, we identify and correct for hot pixels using an extensive set of sky images obtained specifically before and after each visit. As mentioned above \citep[see][]{kevin2022}, we do not carry out flat fielding or dark correction on the spectra. Each science observation is corrected for hot pixels and bias subtracted. Following bias subtraction, we perform cosmic ray correction (taking care of any missing hot pixels) and then the spectral extraction. Therefore, we skip dark and flat field correction. In practice, the subtraction of the dark is achieved by the background subtraction. Figure~\ref{fig:fig5} shows an image and a set of intermediate data products produced by the pipeline. Figure~\ref{fig:all} shows as an example a set of flight images at each reduction step.

\section{Conclusion}\label{sec:conclusion}

We presented here the CONTROL pipeline that is being used to reduce data coming from the CUTE CubeSat mission. Although the pipeline has been designed to reduce data specific to the CUTE mission, its modular approach enables one to employ it also for reducing data obtained by other instruments carrying a long-slit spectrograph and a CCD detector. The main aim of the pipeline is to produce CUTE level 3 and 4 data that are going to be stored at the NASA exoplanet archive and made publicly available starting from 2023. 
The pipeline is available with MIT licence for download at {\tt https://github.com/agsreejith/CONTROL} and it is available also through the CUTE mission webpage.    

The pipeline is being regularly updated following the evolution of the instrument and some new modules will be added in the future if need be. One possible major addition could be that of a module correcting for scattered light that initial observations showed to be present in the data. Measures for mitigating the effect of scattered light have been taken at mechanical design stage by introducing internal light traps \citep{fleming,Egan2020}, but specific in-flight tests are still being run to ensure scattered light does not affect the data quality. 

\acknowledgments
A.G.S. L.F. and M.S. acknowledge financial support from the Austrian Forschungsf\"orderungsgesellschaft FFG project 859718 and 865968. CUTE is supported by NASA grant NNX17AI84G and 80NSSC21K1667 (PI - K. France) to the University of Colorado Boulder. This project was funded in part by the Austrian Science Fund (FWF) [J 4596-N]. We thank the anonymous referee for their comments that helped to significantly improve the paper.

\appendix 

\section{Parameter file} \label{sec:parfile}
The operation of the pipeline  converting level 1 to level 3 data is controlled by an input parameter file that is detailed here below.
\begin{longtable}{p{0.15\textwidth} p{0.8\textwidth}}
data\_path              & Directory indicating the location of the raw data (assumed as current directory if not provided; can be left empty)\\
temp\_path              & Directory indicating where the intermediate files should be stored (default: ``temp'' directory in the raw data directory; the pipeline creates it if not already existing only in case IDL is run with administrator privileges)\\ 
out\_path               & Directory indicating where the output files should be stored (default: ``output'' directory in the raw data directory; the pipeline creates it if not already existing only in case IDL is run with administrator privileges)\\
steps 	                & Reduction steps the pipeline has to execute (described in detail below; default: ``all''; default is assumed if left empty)\\
hb\_map	                & Name and location of hot and bad pixel map (the map has to be of the same size as the CCD and contain either 0, for good pixels, or 1, for bad pixels)\\ 
master\_bias\_file      & Name and location of the master bias file, if available. Set to ``force'' to create a master bias frame with zeros.  (it can be left empty)\\ 
master\_dark\_file      & Name and location of the master dark file, if available. Set to ``force'' to create a master dark frame with zeros.  (it can be left empty)\\
master\_flat\_file      & Name and location of the master flat field file, if available. Set to ``force'' to create a master flat frame with ones. (it can be left empty)\\
save\_temp\_files       & Flag to save intermediate files (options: 0 or 1; default is 1 to save files; default is assumed if left empty)\\
hbpix\_corection\_type  & Type of correction method for hot and bad pixels (options: ``interpolate'' or ``average''; default is ``interpolate''; default is assumed if left empty)\\
bias\_combine\_type     & Type of bias combine (options: ``mean'', ``median'', ``mode''; default is ``median''; default is assumed if left empty)\\
dark\_combine\_type     & Type of dark combine (options: ``mean'', ``median'', ``mode''; default is ``median''; default is assumed if left empty)\\
flat\_combine\_type     & Type of flat combine (options: ``mean'', ``median'', ``mode''; default is ``median''; default is assumed if left empty)\\
saturation\_limit	    & Saturation limit in ADUs of the CCD for data quality check (CUTE specific default are assumed if left empty)\\
sigma\_deviation   	    & Threshold for rejection of pixels when combining the bias and dark calibration files (CUTE specific default are assumed if left empty)\\
cosmic\_ray\_clip  	    & Clipping for the LAcosmic algorithm (CUTE specific default are assumed if left empty)\\
cosmic\_before\_dark    & If set to 1 the cosmic ray correction are executed before dark correction (default is assumed if left empty)\\ 
cent\_poly\_degree 	    & Polynomial degree for the centroid tracing (default is 1)\\
trace\_type	            & Extraction method (options: ``simple'', ``fixed'', ``variable'', ``function''; default: ``simple''; default is assumed if left empty)\\
centroid	            & Centroid value in pixels along the y-axis of the CCD starting from the bottom-left corner (applies to all extraction methods; the pipeline computes it if left empty)\\
slope 	                & Slope of the centroid trace (applies to ``simple'' and ``fixed'' extraction methods; not mandatory; if not provided it uses default value obtained from simulated CUTE data)\\
width 	                & Width of the symmetric spectral extraction window (applies to ``simple'' extraction method; mandatory if ``simple'' extraction method is selected or trace\_type is left empty; in pixels and positive) \\		
upper 	                & Width of the spectral extraction window above the centroid (applies to ``fixed''; mandatory if ``fixed'' extraction method is selected; in pixels and positive)\\    	
lower 	                & Width of the spectral extraction window below the centroid (applies to ``fixed''; mandatory if ``fixed'' extraction method is selected; in pixels and positive)\\ 		
threshold	            & Multiplication factor to the peak value in the cross-dispersion direction defining the limits of the extraction window (applies to ``variable'' and ``function'' extraction methods; mandatory if ``variable'' or ``function'' extraction method is selected; expected to be a number between 0 and 1)\\
background\_trace       & Location of the central part of the background extraction window(s) (it can be positive or negative; it can be more than one number separated by commas; in pixels; if specifying widths please use colon to specify the width of each location, always mandatory)\\
wavecal\_mode	        & Type of wavelength calibration (options: ``crscor'', ``autocor'', and ``simple''; default: ``simple''; default is assumed if left empty; if ``crscor'' is selected the code applies ``simple'' and then carries out cross-correlation with a model file; if ``autocor'' is selected than the code applies ``simple'' and then carries out cross-correlation with the first observation in the set.)\\
wavecal\_file           & Name and location of the wavelength calibration file (i.e., wavelength to pixel map; 1 column file containing wavelengths for each pixel starting from the left side of the detector; mandatory)\\
model\_file	            & Name and location of the stellar model file for cross-correlation (mandatory if ``crscor'' wavelength calibration method is selected)\\
flux\_cal	            & Name and location of the flux calibration file (i.e., count to erg\,s$^{-1}$\,cm$^{-2}$\,\AA$^{-1}$ map; 2 columns file containing wavelengths in \AA\ in the first column and instrument response in the second column; mandatory)\\
\end{longtable}

\subsection{Steps}
The user has the option to select individual steps during reduction. The options are as follows

\begin{tabularx}{\textwidth}{l  X}
hb\_correction & Hot and bad pixel correction\\
create\_mbias  & Create master bias\\
create\_mdark  & Create master dark\\
create\_mflat  & Create master flat\\
cr\_bias       & Remove bias from science and calibration frames\\
cr\_dark	   & Remove dark from science and calibration frames\\
cr\_flat	   & Correct for flat field in science frames\\
cr\_cosmic	   & Correct for cosmic rays in science and calibration frames\\
extract 	   & Define the trace and extract the spectrum\\
bg\_sub	       & Subtract 1D background from the 1D extracted science spectrum\\
wcalib	       & Apply wavelength calibration\\
fluxcalib 	   & Apply flux calibration\\
light\_curve   & Create default light curves from the 1D spectra\\
level3         & Carry out all processes required for generating CUTE Level 3 data (up to flux and wavelength calibration, included). \\
all            & Execute all reduction and calibration steps, up to Level 4 data, included\\
\end{tabularx}

\section{Header}\label{sec:header}

\begin{longtable}{p{0.2\textwidth} p{0.8\textwidth}}
TELESCOP &		Telescope \\
INSTRUME &      Instrument \\
FILENAME &		Filename\\
DATE &          File processing time\\
DATE-OBS &      Date of observation\\
JD		 &		Time of start of observation in Julian date\\
HJD		 &		Time of start of observation in Heliocentric Julian Date\\
MJD		 &		Time of start of observation in Modified Julian Date\\
OBJECT &        Name of the Object observed (from TARGETID) \\    
EQUINX &        Equinox of observational coordinates \\           

RA &            Right Ascension of the target in degrees \\                  
DEC &           Declination of the target in degrees \\                      
RA\_HEX	 &		Right ascension of the target in hexa\\
DEC\_HEX &		Declination of the target in hexa\\
RA\_DEG	 &		Right ascension of the target in degrees\\
DEC\_DEG &		Declination of the target in degrees\\
GLAT	 &		Galactic latitude of the target\\
GLON	 &		Galactic longitude of the target \\
PA		 &		Position angle of the slit in degrees\\ 
GEO\_LAT &		Geocentric latitude\\
GEO\_LON &		Geocentric longitude\\
GEO\_ALT &      Geocentric altitude\\
SUNANGL  &		Sun angle of the observation\\
ERTANGL  &      Earth angle of observation\\
MONANGL	 &		Moon angle of the observation\\
RADANGL &       Radiator angle wrt Sun in degrees \\    
RADERTH &       Radiator angle wrt earth in degrees \\  
SUNERT  &       Sun earth-angle in degrees \\           
AZDIFF  &       Azimuth difference between SUN and CUTE in degrees\\         
APID   &        Packet APID \\                                  
APIDNAME &      Name associated to APID number \\               
FILETYPE &      Type of observation \\                            
FRM\_MID &       ID of this frame in storage \\                    
EXPCMD  &       Commanded exposure duration in ms \\               
EXPTIME &       Measured exposure time in seconds \\                      
TARGETID &      Commanded target ID  \\                            
IMG\_NUM &       Image number within set \\                        
IMG\_TOT &       Total number of images in this set \\             
SHUT\_BEG &      Shutter state at start of image \\                
SHUT\_END &      Shutter state at end of image \\                  
TEC\_VOLT &      TEC Voltage \\                                    
TEC\_CURR &      TEC Current \\                                    
TEC\_HTMP &      TEC Hot Temp \\                                   
CCD\_TEMP &      CCD Temp \\                                       
AVGTEMP &       Mean CCD temperature \\         
MEDTEMP &       Median CCD temperature \\          
STDTEMP &       Standard deviation of CCD temperature \\
RAD\_TEMP &      Radiator Temperature \\                                       
AVGRTEMP &      Mean Radiator temperature \\    
MEDRTEMP &      Median Radiator temperature \\     
STDRTEMP &      Standard deviation of Radiator temperature\\
CCDGAIN &       CCD gain in DN per e- \\                           
RNOISE  &       Readout noise of the CCD \\                        
XB1\_TAI &      XB1 TAI Timestamp \\                              
RA\_AVG  &      Average of RA in degrees \\                 
RA\_STD  &     Standard deviation of RA in degrees \\     
DEC\_AVG &     Average of Declination in degrees \\                
DEC\_STD &     Standard deviation of Declination  in degrees \\    
ROLL\_AVG &    Average of Roll in degrees \\              
ROLL\_STD &    Standard deviation of Roll in degrees \\    
OSX1\_AVG &    Average of left overscan region \\                 
OSX1\_MED &    Median of left overscan region \\                  
OSX1\_STD &    Standard deviation of left overscan region\\                 
OSX2\_AVG &    Average of right overscan region \\                 
OSX2\_MED &    Medain of right overscan region \\                  
OSX2\_STD &    Standard deviation of left overscan region \\    FRM\_ID  &     ID of the frame \\
FRM\_NPK &     Total number of packets in the frame \\
N\_GD\_PKT &   Total number of good packets in the frame \\
N\_FILLED &    Number of missing pixels filled \\
  
YCUT1	 &		Bottom of science box extraction\\	
YCUT2	 &		Top of science box extraction\\

HBPCORR/HBFLG &	Hot and bad pixel correction flag\\
HBMASK		  &	Location of hot and bad pixel mask file\\
HBTYPE		  &	Type of hot and bad pixel correction employed.\\
BIASCOR/BCFLG &	Bias correction flag\\
BIASTYP		  &	Type of bias combine employed\\
BIASSAT		  &	Saturation limit used in bias frames\\
BIASSIG		  &	Deviation limit for good bias frames\\
DARKCOR/DCFLG &	Dark correction flag\\
DARKTYP		  &	Type of Dark combine employed\\
DARKSAT		  &	Saturation limit used in dark frames\\
DARKSIG		  &	Deviation limit for good dark frames\\
FLATCOR/FCFLG &	Flat correction flag\\
FLATTYP		  &	Type of flat combine employed\\
FLATSAT		  &	Saturation limit used in flat frames\\
FLATSIG		  &	Deviation limit for good flat frames\\
CSRYCOR/CRFLG &	Cosmic ray correction flag\\
CRCYCLP		  &	Clip value used for cosmic ray\\
CSRYWEN		  &	Keyword to define when cosmic ray correction was carried out\\
EXTRACT/EXTFLF&	Spectrum extraction flag\\
EXTRTYP		  &	Type of extraction method\\
EXTRCNT		  &	Centroid of extraction spectrum\\
EXTPAR1		  &	Extraction parameter 1\\
EXTPAR2		  &	Extraction parameter 2\\
EXTPAR3		  &	Extraction parameter 3\\
BGSUB/BGFLG   &	Background subtraction flag\\
WAVECAL/WCALFLG&Wavelength calibration flag\\
WCALFLE		 &	Location of Wavelength map file\\
WCALTYP		 &	Type of wavelength calibration employed\\
WLSHFT		 &	Calculated wavelength shift in the data\\
FLUXCAL/FCALFLG &	Flux calibration flag\\
FCALFLE		&   Location of flux response file\\
MEANBIAS	&	Average of bias value subtracted\\
MEANDARK	&   Average of dark value subtracted\\
MEANFLAT	&	Average of flat value divided\\
NFRAMES		&	Number of frames used for creating master files (only for master bias and master dark)\\
NFRAMEMB	&	Number of frames used for creating master bias (in intermediate and calibrated science frames) \\
NFRAMEMD	&	Number of frames used for creating master dark (in intermediate and calibrated science frames)\\
NFRAMEMF	&	Number of frames used for creating master flat (in intermediate and calibrated science frames)\\
NBADPIX      &   Number of bad data quality pixels\\
NBPIXIN & Number of bad data quality pixels falling in the science and background extraction boxes\\
DQFACTR		&	Number of bad data quality pixels as a fraction of total number of pixels\\
PIPENUM & Pipeline version number used to reduce the data\\
TFIELDS	&		Number of fields in each row\\
TTYPEn  &		Type of the content in the field\\
TFORMn  &		Format of the content in the field\\
TUNITn  &		Unit of the content in the field\\
TDISPn  &		Display format for the content in the field\\
TNULLn  &		Undefined value for the content in the field\\

\end{longtable}

\section{Dependencies} \label{sec:deps}
The software requires previous installation of the IDL astronomy library\footnote{http://idlastro.gsfc.nasa.gov/}, Coyote IDL library\footnote{http://www.idlcoyote.com/documents/programs.php}, MPFIT functions from Markwardt IDL Library\footnote{https://pages.physics.wisc.edu/~craigm/idl/fitting.html}, and LA cosmic for cosmic ray correction\footnote{http://www.astro.yale.edu/dokkum/lacosmic/}. Other codes required by the program are provided with the software distribution.

\bibliography{report}{}

\end{document}